\documentclass[aps, prx, reprint, superscriptaddress, longbibliography, floatfix,%reprint,
tightenlines,
nofootinbib,
 amsmath,amssymb,
]
{revtex4-2}
\usepackage{graphicx}% Include figure files
\usepackage{dcolumn}% Align table columns on decimal point
\usepackage{bm}% bold math
\usepackage{physics}
\usepackage{soul}
\DeclareMathAlphabet{\pazocal}{OMS}{zplm}{m}{n}

\usepackage{amsmath}
\usepackage{amssymb}
\usepackage{comment}
\usepackage{stmaryrd}
\usepackage{braket}
\usepackage{amsfonts}
\usepackage{mathrsfs}
\usepackage{subfigure}
\usepackage{xcolor}
\usepackage[colorlinks=true,linkcolor=blue,anchorcolor=red,citecolor=blue, urlcolor=blue]{hyperref}
\begin{document}
\title{Neural Network Representation of Generalized Parton Distributions (NNGPD)}

\author{Jitao Xu}
\thanks{These authors contributed equally to this work.}
\affiliation{CS Department, Old Dominion University, Norfolk, VA 22904, USA.}

\author{Ho Jang}
\thanks{These authors contributed equally to this work.}
%\email{sgv2ew@virginia.edu}
\affiliation{Department of Physics, University of Virginia, Charlottesville, VA 22904, USA.}

\author{Zaki Panjsheeri}
\thanks{These authors contributed equally to this work.}
%\email{zap2nd@virginia.edu}
\affiliation{Department of Physics, University of Virginia, Charlottesville, VA 22904, USA.}

\author{Gia-Wei~Chern}
\thanks{Corresponding author}
\email{gchern@virginia.edu}
\affiliation{Department of Physics, University of Virginia, Charlottesville, VA 22904, USA.}

\author{Yaohang~Li}
\thanks{Corresponding author}
\email{yaohang@cs.odu.edu}
\affiliation{CS Department, Old Dominion University, Norfolk, VA 22904, USA.}

\author{Simonetta Liuti} 
\thanks{Corresponding author}
\email{sl4y@virginia.edu}
\affiliation{Department of Physics, University of Virginia, Charlottesville, VA 22904, USA.}

\author{Douglas Q. Adams} 
%\email{yax6jr@virginia.edu}
\affiliation{Department of Physics, University of Virginia, Charlottesville, VA 22904, USA.}

\author{Michael~Engelhardt} 
%\email{engel@nmsu.edu}
\affiliation{Department of Physics, New Mexico State University, Las Cruces, NM 88003, USA}
 
\author {Gary R. Goldstein}
\affiliation{Department of Phyiscs, Tufts University, Medford, MA 02155}

\author{Adil Khawaja} 
%\email{taz8sp@virginia.edu}
\affiliation{Department of Physics, University of Virginia, Charlottesville, VA 22904, USA.}

\author{Huey-Wen Lin}
\affiliation{Department of Physics and Astronomy, Michigan State University, East Lansing, MI , 48824 USA.}

\author{Saraswati Pandey} 
%\email{qbs8mh@virginia.edu}
\affiliation{Department of Physics, University of Virginia, Charlottesville, VA 22904, USA.}

\author{Kemal~Tezgin}
\affiliation{Department of Physics, Virginia Tech, Blacksburg, VA 24061, USA.}

\begin{abstract}
\vspace{0.2cm}
\begin{center}
\large
    EXCLAIM Collaboration
    %%%
    \normalsize
\end{center}

We present a neural-network–based framework for modeling generalized parton distributions, referred to as NNGPD, in which GPDs are represented as flexible functions constrained through physically motivated integral relations. In this approach, experimental and theoretical information is incorporated into the training procedure via loss functions enforcing convolution integrals that define Compton form factors, as well as Mellin moments related to generalized form factors accessible in lattice QCD. This formulation reflects the inverse-problem character of GPD phenomenology without assuming a specific functional ansatz. As a proof of concept, we benchmark the NNGPD framework using a phenomenological spectator-based GPD model, from which synthetic training data for Compton form factors and Mellin moments are generated. The neural network is trained solely on these aggregate observables, and the resulting GPDs are compared directly with the underlying model distributions in a closure-type test. We find that the neural-network representation reproduces the main features of the GPDs over the relevant kinematic domain, despite being constrained only by their integral projections. This study demonstrates the viability of neural-network representations of GPDs constrained by global physical observables and provides a basis for future phenomenological applications combining experimental measurements of deeply virtual Compton scattering, including those anticipated at the Electron–Ion Collider, with lattice QCD inputs for Mellin moments and generalized form factors.
\end{abstract}

\date{\today}
\maketitle

\section{Introduction}

\label{sec:intro}

The extraction of the QCD parton correlation functions from experiment is intrinsically an inverse problem in that they are not directly observable but they enter the physical processes through integral relations involving known kernels. 
Specific projections are given, for instance, by the parton distribution functions (PDFs) from deep inelastic scattering, the transverse momentum distributions (TMDs), and the  
generalized parton distributions (GPDs). 
Machine-learning (ML) approaches have proven highly effective in addressing inverse problems of this type. For example, the PDF and TMD communities have pioneered artificial neural network (ANN)–based frameworks aimed at reducing the functional-form bias inherent in traditional parametric fits  \cite{NNPDF:2014otw, Carrazza:2019mzf, NNPDF:2021njg, NNPDF:2024dpb, Bacchetta:2025ara}.
The inverse problem associated with GPDs is considerably more stringent than for other parton distributions, as the dependence on the parton longitudinal momentum fraction 
$x$ cannot be accessed directly in experiment. Instead, it is integrated over in the measured observables, which are expressed in terms of Compton form factors (CFFs) probed in deeply virtual Compton scattering (DVCS) and related hard exclusive processes. 
To extract GPDs from DVCS-type data, ML approaches become, therefore, indispensable. More recently, sophisticated deep-learning methods have been developed to extract CFFs directly from DVCS data~\cite{Almaeen:2024guo,Almaeen:2022imx,hossen:2024}, and, more generally, to uncover underlying functional structures in a data-driven yet interpretable manner~\cite{Dotson:2025omi}. Despite these advances, the extraction of GPDs remains substantially more challenging than that of other QCD correlations, owing to their higher dimensionality.

Another source of information on GPDs is from lattice QCD (LQCD), where the Mellin moments have been evaluated up to $n=5$ for the proton \cite{Bhattacharya:2023ays}. %, and recent calculations are attempting now to achieve $n = 6$ for the pion \cite{Shindler:2023xpd}. 
Consequently, any phenomenologically viable representation of GPDs must satisfy a hierarchy of global integral constraints (for reviews on GPDs we refer the reader to Refs. \cite{Diehl:2003ny, Belitsky:2005qn, Kumericki:2016ehc}).

%~\cite{NNPDF:2014otw,Carrazza:2019mzf,NNPDF:2021njg,NNPDF:2024dpb,Bacchetta:2025ara}. 

These challenges motivate learning frameworks that go beyond purely data-driven fitting and explicitly incorporate the theoretical structure of the GPD inverse problem. A promising direction is provided by ML approaches in which global or integral constraints are imposed directly at the level of the loss function, thereby guiding optimization toward physically admissible solutions. In this setting, experimental observables such as CFFs and theoretical inputs such as Mellin moments, polynomiality conditions, and sum rules enter naturally as integral constraints that couple the learned GPDs across momentum fractions and kinematic variables. This philosophy is closely related to physics-informed neural networks (PINNs), originally developed to solve differential equations by embedding governing equations into the training objective~\cite{Raissi2018,Raissi2019,Karniadakis2021}. In PINNs, physical information typically enters through loss terms involving differential operators acting on the neural-network output.

For GPD extraction, however, the relevant constraints are inherently nonlocal and integral in nature: experimental and lattice observables constrain integral transforms and moments of the underlying distributions rather than their local values. While this structure differs from the local differential constraints typically employed in PINNs, it can be naturally incorporated within the same loss-function–based framework. In this sense, integral-constrained learning provides a generalization of physics-informed neural networks, with convolution kernels and moment relations replacing differential operators as the primary carriers of physical information.

In this work, we adopt this constrained-learning strategy to construct a neural-network representation of GPDs. 
%The network is trained to reproduce CFFs and lattice-QCD Mellin moments by enforcing the corresponding integral relations through the loss function. 
As a proof of concept, we benchmark the framework using a flexible, Reggeized spectator model-based parameterization of the GPD
~\cite{Goldstein:2011, Kriesten:2021sqc,Panjsheeri:2025vpa}, which provides controlled input for simulations of form-factor and moment data. In particular, we generate training data for the NNGPD framework using the latest iteration of this parameterization, UVA2 \cite{Panjsheeri:2025vpa}.  By construction, the resulting neural-network parametrization satisfies all imposed sum rules and moment constraints within the quoted uncertainties, while remaining agnostic about the detailed functional form of the GPDs. This approach therefore provides a systematic and data-driven framework for GPD phenomenology, unifying experimental measurements, lattice calculations, and theoretical constraints within a single machine-learning setting.
In this respect, this is a closure test  of the GPD extraction, or a first step which is used to check whether the NNGPD proposed fitting procedure can recover the underlying distributions. It serves as a necessary precursor to the implementation of both CFF and LQCD data upon their availability.
%%%%
\footnote{Existing measurements are currently not sufficient to provide realistic constraints.}
%%%%

The remainder of this paper is organized as follows. Section~\ref{sec:GPD} reviews the definition and key properties of GPDs, emphasizing their connection to DVCS and future measurements at the Electron–Ion Collider (EIC). In Sec.~\ref{sec:nn-model}, we introduce the neural-network framework and we describe how global integral constraints from CFFs, electromagnetic form factors, and Mellin moments are implemented through the loss function. 
Section~\ref{sec:results} presents a proof-of-principle study based on the UVA2 parameterization, using synthetic data to benchmark the reconstruction accuracy of the neural network in a closure test.
Section~\ref{sec:conclusion} concludes with a summary and an outlook towards the implementation of polynomiality and other properties as hard constraints, while incorporating experimental and LQCD inputs in the loss.

%%%%%%%%%%
%%%%%%%%%%  SECTION II
%%%%%%%%%%
\section{Generalized Parton distributions and DVCS Observables}
\label{sec:GPD}

GPDs provide a unified framework for describing the longitudinal momentum and transverse spatial structure of hadrons, 
encoding rich information about correlations between momentum, position, spin, and orbital angular momentum degrees of freedom inside the proton.
%interpolating between ordinary parton distribution functions (PDFs) and elastic form factors. 
They arise naturally in the parametrization of nonperturbative quark and gluon correlations probed in deeply virtual exclusive processes (DVES), such as DVCS and deeply virtual meson production (DVMP) in a QCD factorized scenario \cite{Muller:1994ses, Ji:1996ek, Radyushkin:1997ki}. 
%As such, GPDs encode rich information about hadron structure, including correlations between momentum, position, spin, and orbital angular momentum degrees of freedom.

%Let us denote by $p$ ($p^\prime$) and $\Lambda$ ($\Lambda^\prime$) the incoming (outgoing) nucleon momentum and helicity, respectively. 
GPDs are defined through off-forward nucleon matrix elements of bilocal quark and gluon operators. For quark GPDs, these matrix elements take the general form, at a given renormalization scale, $\mu^2$,
\footnote{Throughout this work, we choose $\mu^2 = Q^2$ for notational simplicity and consistency with the CFF sector.} 
%Apart from the $n=0$ case, the Mellin moments and hence the respective GFFs, also depend on the renormalization scale $\mu$.
\begin{eqnarray}
\label{matrix-element-GPDs}
&&W^\Gamma_{\Lambda \Lambda'} =  P^{+}\int\frac{dz^{-}}{2\pi}e^{ixP^{+}z^{-}}  \\
   && \quad \,\,\, \bra{p',\Lambda'}\bar\psi_q\Bigl(-\frac{z}{2} \Bigr)\mathcal{W}\Bigl(-\frac{z}{2},\frac{z}{2} \Bigr)\,\Gamma \psi_q\Bigl(\frac{z}{2} \Bigr)\ket{p,\Lambda}\bigg|_{\stackrel{\displaystyle{z^{+}=0}}{\displaystyle{\vec{z}_T=0}}}, \nonumber 
\end{eqnarray}
where $\Gamma$ specifies the Dirac structure of the operator; $P = (p+p')/2$ is the average nucleon momentum, $x$ denotes the average longitudinal momentum fraction carried by the active parton, $\Delta=p-p'$, is the momentum transfer between the incoming and outgoing protons, that introduces two additional invariants, $t=\Delta^2$, the invariant momentum transfer squared, and  the skewness parameter $\xi = -\Delta^+/(2P^+)$, or the fraction of the longitudinal momentum transfer carried by the quarks. The separation between the quark fields is chosen along the light cone, $z^2=0$, ensuring sensitivity to leading-twist partonic dynamics. Gauge invariance of the nonlocal operator is guaranteed by the Wilson line
\begin{equation}
    \mathcal{W}(a,b) = \mathcal{P}\,\mbox{exp}\Bigl(ig\int_b^a dx_\mu\,A^\mu\Bigr)\, ,
\end{equation}
which resums gluon exchanges along the integration path connecting the two quark fields. Here $\mathcal{P}$ denotes path ordering, and $A^\mu$ is the gauge field. In light-cone gauge, this Wilson line formally reduces to unity.

The specific choice of $\Gamma$ determines both the symmetry properties and the twist classification of the correlator. In this work, although the neural-network methodology we develop is generic and does not depend on a particular Dirac structure, we focus on the reconstruction of leading-twist quark GPDs obtained from Mellin moments in the vector sector  $\Gamma = \gamma^+$, which conserves the helicity of the active quark and defines the chiral-even, unpolarized GPDs. $W_{\Lambda \Lambda'}^{\gamma^+}$ can be parametrized in terms of two GPDs, corresponding to the proton helicity conserving $(\Lambda=\Lambda')$, $H$, and helicity flip $(\Lambda \neq \Lambda')$, $E$, cases,  
%%%%
\begin{eqnarray}
\label{chiral-even-GPDs}
  \!\!  W_{\Lambda \Lambda'}^{\gamma^+} \!
   \! &=& \!
    \bar u_{\Lambda'}(p')\Big[H^q(x, \xi, t)\,\gamma^+
    + E^q(x, \xi, t) \, \frac{i\sigma^{+\alpha}\Delta_\alpha}{2M}\bigg] u_\Lambda(p)\, \nonumber \\
\end{eqnarray}
where $M$ denotes the nucleon mass.
%where $\Delta = p' - p$ is the momentum transfer$. 

The functions $H^q(x,\xi,t)$ and $E^q(x,\xi,t)$ are the unpolarized quark GPDs. In the forward limit $\xi \to 0$ and $t \to 0$, $H_q$ reduces to the familiar quark PDF, $q(x)$. The GPD $E_{q}$, however, has no PDF equivalent in the forward limit.  
%and Pauli form factors, respectively.

%A second class of leading-twist, helicity-conserving quark GPDs is obtained by choosing the axial-vector Dirac structure $\Gamma = \gamma^+\gamma_5$. This choice probes the spin-dependent partonic structure of the nucleon and leads to the parametrization
%\begin{widetext}
%\begin{align}\label{evenGPDsflip}
%    P^{+}\int\frac{dz^{-}}{2\pi}e^{ixP^{+}z^{-}} & \bra{p',\lambda'}\bar\psi_q\Bigl(-\frac{z}{2}\Bigr)\mathcal{W}\Bigl(-\frac{z}{2},\frac{z}{2}\Bigr)\,\gamma^+\gamma_5\psi_q\Bigl(\frac{z}{2}\Bigr)\ket{p,\lambda}\bigg|_{z^{+}=0,\vec{z}_T=0}\nonumber\\
%    &=\bar u(p',\lambda')\bigg[\tilde H^q(x, \xi, t)\,\gamma^+\gamma_5 + \tilde E^q(x, \xi, t)\,\frac{\gamma_5\Delta^+}{2m}\bigg] u(p,\lambda)\, .
%\end{align}
%\end{widetext}
%The axial GPDs $\tilde H^q$ and $\tilde E^q$ encode information about the helicity and pseudoscalar structure of the nucleon and are directly related to polarized PDFs and axial form factors in appropriate kinematic limits.
%Beyond their formal role as nonperturbative QCD correlation functions, GPDs are of central importance for the experimental exploration of hadron structure. 

%Through these processes, GPDs offer access to fundamental questions such as the spatial distribution of quarks and gluons, correlations between parton momentum and transverse position, and the decomposition of the nucleon spin into its intrinsic and orbital components.

\subsection{Sum Rules and Polynomiality}
%%%%%%%%
%%
The $n=1$ Mellin moments of the vector GPDs give the quark, $q$, contribution to the Dirac and Pauli form factors, 
%The property of polynomiality restricts the moments to be polynomials in $\xi$ with coefficients that are a function of $t$ alone. The coefficients are known as Generalized Form Factors (GFFs).
%are independent of $\xi$ and reduce to the electromagnetic form factors,
\begin{subequations}
\label{eq:DiracPauli}
\begin{eqnarray}
\int_{-1}^{1} dx\, H^q(x,\xi,t) &=& F_1^q(t), \\
\int_{-1}^{1} dx\, E^q(x,\xi,t) &=& F_2^q(t),
\end{eqnarray}
\end{subequations}
%where $F_1^q(t)$ and $F_2^q(t)$ are the Dirac and Pauli form factors, respectively, 
which are experimentally accessible in elastic electron scattering from both proton and neutron targets \cite{Cates:2011pz, Qattan:2012zf, Qattan:2024pco}.

The second Mellin moments ($n=2$) read,
\begin{subequations}
\label{eq:2ndmoment}
\begin{eqnarray}
\int_{-1}^{1} dx\, x\, H^q(x,\xi,t) &=& A_q(t) + (2\xi)^2\, C_q(t), \\
\int_{-1}^{1} dx\, x\, E^q(x,\xi,t) &=& B_q(t) - (2\xi)^2\, C_q(t),
\end{eqnarray}
\end{subequations}
where $A_q(t)$, and $B_q(t)$, remarkably, parametrize the matrix element between proton states of the QCD energy--momentum tensor, and give origin to the Ji sum rule for the proton angular momentum \cite{Ji:1996ek},
\begin{equation}
    \frac{1}{2} = J^q + J^g , \quad\quad J^{q,g} = \frac{1}{2} \Big[ A_{q,g}(0)+ B_{q,g} (0) \Big] 
\end{equation}
where $A_{q,g}(t)$, $B_{q, g}(t)$, and $C_{q, g}(t)$ are the defined as the order $n=2$, quark (gluon) generalized form factors (GFFs). 
%%%%%
%\footnote{The GFF $C_{q,g} = $ 
The form factor $A_q(t)$ encodes the momentum fraction carried by quarks, $B_q(t)$ is related to the total angular momentum through Ji's sum rule, and 
$C_q(t)$ (often referred to as the $D$-term form factor) is believed to describe the shear-force distributions inside the nucleon \cite{Polyakov:2002yz, Polyakov:2018zvc}. 

Higher Mellin moments similarly generate towers of GFFs associated with higher-spin local operators.
\begin{equation}
    M_n^q(\xi, t) \equiv \int_{-1}^{1} dx x^{n-1} F_q(x, \xi, t), \quad\quad F=H,E.
\end{equation}
One can show that the $x^n$ moments of the quark GPDs are even polynomials in $\xi$, namely,
\begin{eqnarray}
\label{eq:H_moment}
    \int_{-1}^{1} dx x^{n-1} H_q(x, \xi, t) & = & \sum_{i=0,even}^{n-1} (2\xi)^i A^q_{n,i}(t) \\
    &&+ \mod(n-1,2) (2\xi)^{n} C_{n}^q(t) \nonumber \\
\label{eq:E_moment}
    \int_{-1}^{1} dx x^{n-1} E_q(x, \xi, t) & = & \sum_{i=0,even}^{n-1} (2\xi)^i B^q_{n,i}(t) \\
    &&- \mod(n-1,2) (2\xi)^{n} C_{n}^q(t) . \nonumber 
\end{eqnarray}
These expressions define the polynomiality property of GPDs. 

%Due to polynomiality, $H^n$ has the following form for $n = 0,1$ and $2$,
%
%%
%\begin{eqnarray}
%H^0(\xi, t) &=& A_{10}(t) \\
%
%H^1(\xi, t) &=& A_{20}(t) + (2\xi)^2 C_{20}(t) \\
%
%H^2(\xi, t) &=& A_{30}(t) + (2\xi)^2 A_{32}(t)
%\end{eqnarray}
%%
%
%For $\xi = 0$, one can treat the moments in a fashion similar to what is done for PDFs and use the symmetry property $\bar{q}(x) = q(-x)$ to obtain an expression similar to eq.(\ref{eq:pdfMomOPE}) ,
%
%%
%\begin{equation}
%    H^n_q(0,t) = \int_0^1 dx x^n (H_q(x, 0, t) + (-1)^{n+1}H_{\bar{q}}(x, 0, t))
%\end{equation}
%%
%We next consider the constraints arising from Mellin moments of GPDs. The $n$-th Mellin moment of a quark GPD $F_q$ is formally defined as
%\begin{eqnarray}
%M_n^q(\xi,t;\mu^2) = \int_{-1}^{1} dx\, x^{n-1}\, F_q(x,\xi,t;\mu^2),
%\end{eqnarray}
%where $\mu^2$ denotes the renormalization scale. Throughout this work, we choose $\mu^2 = Q^2$ for notational simplicity and consistency with the CFF sector.
%Unlike CFFs, Mellin moments correspond to matrix elements of local twist-two operators and therefore admit a polynomial expansion in $\xi$, known as the polynomiality property. The coefficients of this expansion are generalized form factors that depend only on $t$ and the renormalization scale, providing particularly clean and theoretically controlled constraints on GPDs.

%%%%% SUBSEC IIB
%%%%%
\subsection{DVES Observables: Compton form factors}
\label{sec:DVES}

%%%%%%%%%%%%
\begin{figure}
\centering
\includegraphics[width=0.7\columnwidth]{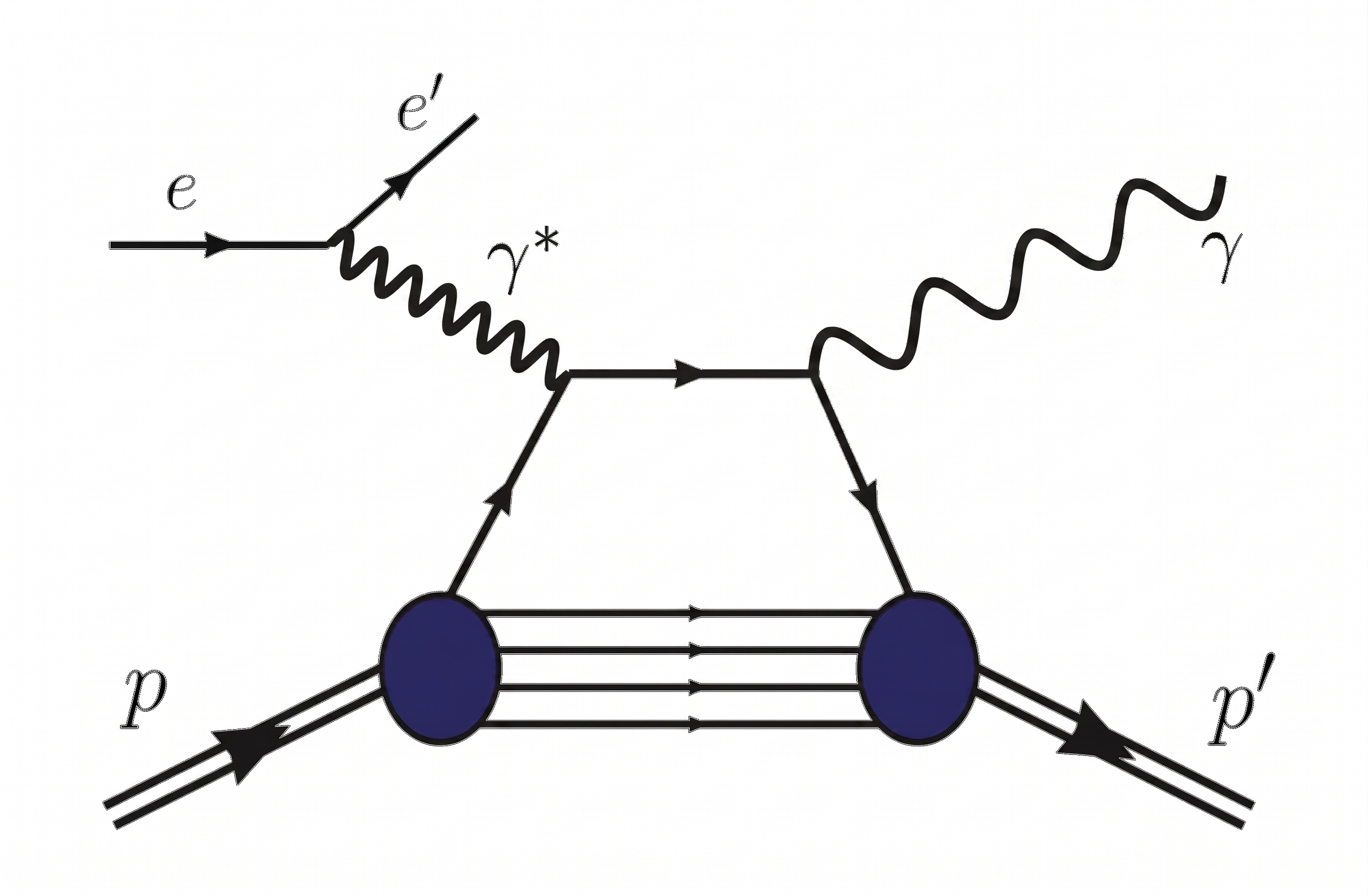}
\caption{Tree-level GPD Feynman diagram for the DVCS amplitude, $ep \rightarrow e'p' \gamma$.}
    \label{fig:DVES}
\end{figure}
%%%%%%%%%%
%%%
GPDs enter the observables for the DVES processes (Figure \ref{fig:DVES}) via convolution integrals with  hard-scattering kernels perturbatively calculable in QCD, leading to CFFs. 
%The extraction of GPDs from experimental data is therefore inherently an inverse problem, complicated by limited kinematic coverage and correlations among multiple observables. This challenge has motivated the development of flexible and theoretically consistent parametrizations, capable of incorporating diverse experimental constraints while maintaining fidelity to the underlying QCD structure.
%
Focusing on DVCS, where the outgoing particle at the hard scattering vertex is a photon, one has that 
at leading twist, and leading order in perturbative QCD (PQCD), the CFF, ${\mathcal F}_q$, associated with a quark GPD, $F_q$, is given by a convolution integral,
\begin{equation}
\label{eq:CFF}
\mathcal{F}_q(\xi,t,Q^2)
=
\int_{-1}^{1} dx\,
C^{(\pm)}(x,\xi)\,
F_q(x,\xi,t;Q^2),
\end{equation}
where $C^{(\pm)}(x,\xi)$ is the Wilson coefficient function at lowest order. Explicitly,
\begin{eqnarray}
 \label{eq:WCF}
 { C^\pm(x,\xi) = \frac{1}{x-\xi + i \epsilon} \pm \frac{1}{x+\xi - i \epsilon}  } , %C^\pm(x,\xi) = \frac{1}{x-\xi - i \epsilon} \pm \frac{1}{x+\xi - i \epsilon},
\end{eqnarray}
The real and imaginary parts of the CFFs can be separated out as,
\begin{eqnarray}
    \mathcal{F}_q(\xi,t)  &=& \Re  \mathcal{F}_q + i \, \Im \mathcal{F}_q\\
    &=& P.V. \int_{-1}^{1} dx  \left( \frac{1}{x-\xi} + \frac{1}{x+\xi} \right)  F_q(x,\xi,t) \nonumber \\
    && {- i \pi \, \big(F_q(\xi,\xi,t) - F_q(-\xi,\xi,t)\big)}\, . \nonumber
\end{eqnarray}
%Here the antisymmetric kernel ($-$) applies to the vector GPDs $H$ and $E$, while the symmetric kernel ($+$) applies to the axial-vector GPDs $\widetilde H$ and $\widetilde E$. 
The hard-scattering kernel encodes the short-distance dynamics of the DVCS subprocess and introduces a nontrivial analytic structure, such that the resulting CFF is generally complex-valued. Physically, the imaginary part of a CFF probes the underlying GPD along the cross-over line $x=\pm\xi$, whereas the real part involves a principal-value integral over the full $x$ domain. 
%As a consequence, CFF measurements impose intrinsically nonlocal constraints on the underlying GPDs.

%%%%%
%%%%%
%%%%%
\subsection{Symmetry and Endpoint Constraints}
\label{sec:symmetry}
In addition to global integral relations, GPDs satisfy exact symmetry properties and boundary conditions. These constraints are independent of phenomenological modeling and therefore provide physics-driven inputs that can be incorporated through  specific terms in the loss function. 
%In this subsection, we summarize the physical origin of the symmetry constraints under $x\to -x$, and the endpoint conditions imposed on the leading-twist vector GPDs $H$ and $E$.

The symmetry properties under reversal of the longitudinal momentum fraction, $x\to -x$, originate from charge conjugation.
%the negative values of $x$ corresponding to antiquark degrees of freedom. 
For quark GPDs, the region $x>0$ corresponds to quark emission and absorption, while $x<0$ describes antiquark contributions. The charge-conjugation symmetry relates these two regions and leads to definite properties under $x\to -x$.
%which depend on the Dirac structure of the underlying bilocal quark operator.
Specifically, we define the unpolarized anti-quark GPD, $H_{\bar{q}}(x,\xi,t)$, $E_{\bar{q}}(x,\xi,t)$ as, 
\begin{subequations}
\begin{eqnarray}
    H_{\bar{q}}(x,\xi,t) = - H_{{q}}(-x,\xi,t) \\
     E_{\bar{q}}(x,\xi,t) = - E_{{q}}(-x,\xi,t)
\end{eqnarray}
\end{subequations}
The $``+"$ and $``-"$ contributions are defined as, 
\begin{subequations}
\label{eq:Hplusminus}
\begin{eqnarray}
    H^+_q(x,\xi,t) = H_{{q}}(x,\xi,t)  + H_{\bar{q}}(x,\xi,t) \\
    H^-_q(x,\xi,t) = H_{{q}}(x,\xi,t)  - H_{\bar{q}}(x,\xi,t)
\end{eqnarray}
\end{subequations}
with the following symmetry properties for $x \rightarrow -x$ (we omit writing the dependence on the $\xi,t$ variables, which are not symmetry arguments),
\begin{subequations}
\label{eq:Hplusminu}
\begin{eqnarray}
H^-_q(-x) &= & H^-_q(x) \quad\quad symmetric \; \\
H^+_q(-x) &=& -H^+_q(x) \quad antisymmetric 
 \end{eqnarray}
\end{subequations}
We reiterate that these symmetry properties are exact consequences of the transformation behavior of the quark bilinears in the vector sector under charge conjugation and must be satisfied by any physically admissible GPD representation. Enforcing these relations reduces the functional freedom of the parameterization and ensures consistency between quark and antiquark contributions across the entire support region.

In addition to these symmetry constraints, GPDs obey nontrivial boundary conditions at the kinematic endpoints $x=\pm 1$, where,
%The endpoint $x=1$ corresponds to the unphysical configuration in which a single parton carries the entire longitudinal momentum of the hadron, leaving all spectator degrees of freedom with vanishing momentum. For any normalizable light-cone wave function, such configurations have zero measure. As a result, leading-twist vector GPDs vanish at the endpoints,
\begin{equation}
H_q^\pm(x=\pm 1,\xi,t) = 0, 
\qquad 
E_q^\pm(x=\pm 1,\xi,t) = 0.
\end{equation}
These conditions represent the natural generalization of the familiar endpoint behavior of ordinary parton distribution functions in the forward limit.
Note that for  the antisymmetric, $``+"$ combination, because GPDs are continuous functions of $x$, another symmetry constraint follows, given by, $H^+(0,\xi,t)=E^+(0,\xi,t)=0$.  

Both the symmetry and endpoint constraints play an essential role in stabilizing the global integral relations discussed above. Violations of these symmetry relations, as well as non-vanishing endpoint values would generate spurious contributions to Mellin moments and may spoil their required polynomial dependence on $\xi$. Similarly, endpoint pathologies can lead to numerical instabilities in the convolution integrals defining CFFs. Enforcing these constraints therefore restricts the learning problem to a physically admissible function space consistent with fundamental field-theoretic principles.

\begin{figure*}
\centering
\includegraphics[width=1.75\columnwidth]{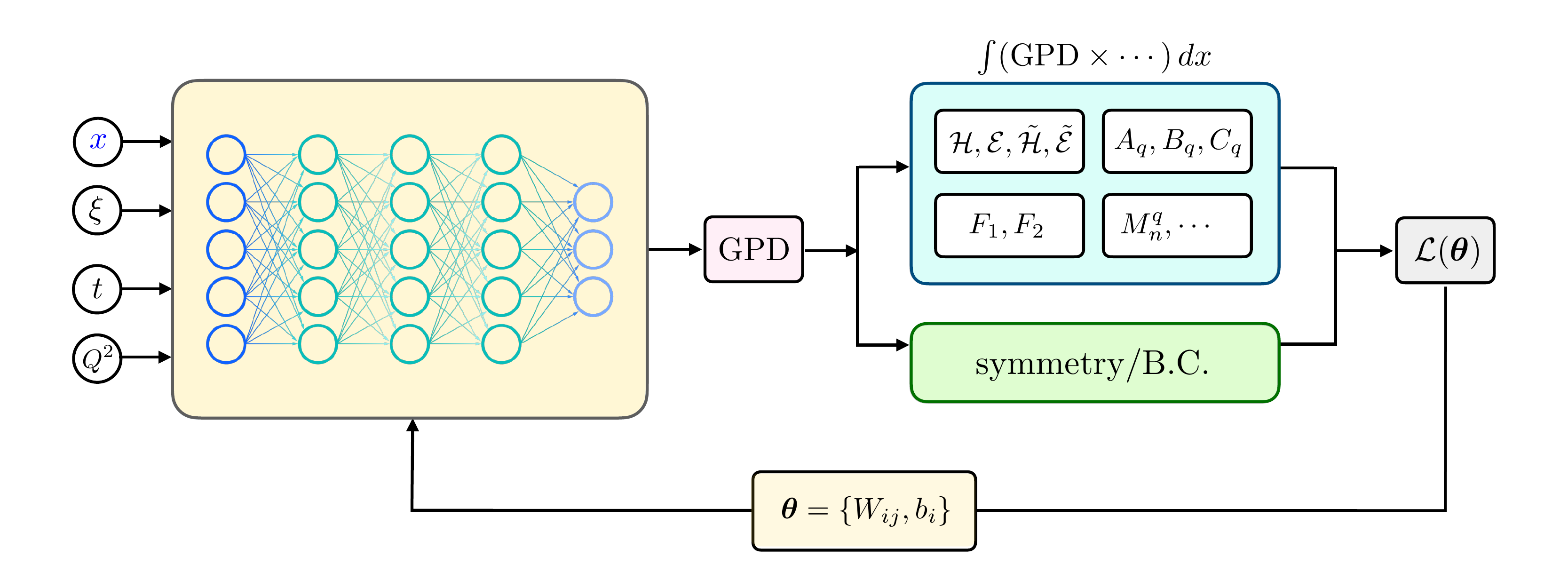}
\caption{Physics-informed machine-learning framework for generalized parton distributions (GPDs).
A neural network maps the kinematic variables $(x,\xi,t,Q^2)$ to the leading-twist GPDs $H$, $E$, $\widetilde H$, and $\widetilde E$. The network is trained using a composite loss function encoding physical constraints, including integral relations for Compton form factors and Mellin moments, whose special limits reproduce electromagnetic and generalized gravitational form factors. Additional loss terms enforce symmetry under $x \to -x$ and boundary conditions such as vanishing at $x=1$. Constraints are provided by experimental measurements of hard exclusive processes and lattice QCD calculations, enabling a unified, data-driven extraction of GPDs.}
    \label{fig:gpd-ml-schematic}
\end{figure*}

\section{Neural network model}
\label{sec:nn-model}
%%%%
%As discussed in Sec.~\ref{sec:intro}, the determination of GPDs from experimental and theoretical information is intrinsically a \emph{global inverse problem}. GPDs are not directly observable quantities; instead, they enter physical processes through convolution integrals with known hard-scattering kernels, such as those appearing in DVCS, and through their Mellin moments. Consequently, any phenomenologically viable representation of GPDs must satisfy a hierarchy of integral relations, symmetry constraints, and boundary conditions, rather than a collection of pointwise constraints in kinematic space.

%Traditional phenomenological approaches have relied on physics-motivated functional ans\"atze that implement these constraints analytically. While such parametrizations have proven successful, they inevitably limit the functional flexibility of GPDs and may introduce biases when confronted with increasingly precise and heterogeneous data sets. 

In this Section, we introduce a physics-informed deep-learning framework for GPDs that is designed to incorporate theoretical structure and experimental information in a unified and systematic way. In a nutshell, we develop an approach where, instead of prescribing a specific functional form, one can represent GPDs using highly expressive neural networks whose parameters are determined solely by enforcing known physical laws and experimentally accessible observables.

%The central idea is 
%to represent the GPDs, $H$ and $E$, as neural-network functions of their natural kinematic variables $(x,\xi,t,Q^2)$, while constraining the learning process through loss functions constructed from integral relations, symmetry properties, and boundary conditions. This approach allows 
%to field-theoretic knowledge—such as sum rules, polynomiality, and crossing symmetries—to be embedded directly at the level of optimization, without imposing restrictive analytic parametrizations.

Figure~\ref{fig:gpd-ml-schematic} provides a schematic overview of the proposed framework. A neural network, parameterized by weights and biases $\bm\theta = \{W_{ij}, b_i\}$, takes the kinematic variables $(x,\xi,t,Q^2)$ as inputs and outputs, the GPDs $H_q$, and $E_q$ 
%%%%%
\footnote{The same scheme can be extended to include the  GPDs in the axial vector sector, $\widetilde H$, and $\widetilde E$, as well as the chiral-odd GPDs, which, however, we do not consider in the present paper.}
%%%%%
%%
The determination of the network parameters is entirely driven by a composite loss function, which includes contributions from integral constraints associated with: the CFFs, ${\cal H}...$, Eq.\eqref{eq:CFF}, the $n=1$ Mellin moments given by the Dirac and Pauli form factors, $F_1^q$, $F_2^q$, Eqs.\eqref{eq:DiracPauli}, as well as the $n=2$  $A_q(t)$, $B_q(t)$, $C_q(t)$, Eqs.\eqref{eq:2ndmoment}, and higher order, up to $n=7$, Mellin moments. Additional loss terms enforce symmetry properties of GPDs under $x \to -x$ and boundary conditions such as $H_q(x=1,\xi,t,Q^2)=0$.
In what follows, for simplicity,  we demonstrate our procedure on the GPD $H_q$, with data for the $u$-quark only. Our method can be easily extended to all other GPDs and flavors. 

Within this framework, the training data naturally combine inputs from multiple complementary sources. 
%Experimental measurements of DVCS and related hard exclusive processes do not access GPDs directly, but instead constrain the associated CFFs through convolution integrals over the longitudinal momentum fraction $x$. 
%In parallel, lattice QCD calculations provide access to Mellin moments of GPDs at discrete values of the momentum transfer $t$ and the renormalization scale $Q^2$. 
By unifying the heterogeneous constraints from experiment, LQCD and the symmetries of the theory within a single optimization problem, the framework recasts the extraction of GPDs as a {\it constrained learning task governed by global integral relations rather than pointwise supervision}. This constitutes our point of departure from standard PINN approaches, and the novelty introduced by our approach.
%This formulation provides a flexible and systematically improvable pathway toward nonperturbative hadron structure. 
In the following, we discuss the individual contributions to the total loss function, beginning with the constraint imposed by Compton form factors.

\subsection*{Compton Form Factor Loss}

Given a set of experimentally extracted CFF values at discrete kinematic points $\{t_m,\xi_m,Q_m^2\}$, the corresponding contribution to the loss function is constructed by comparing these data to the CFFs computed from the ML-predicted GPDs. Schematically, the CFF loss can be written as
\begin{eqnarray}
	\label{eq:L_CFF}
	& & \mathcal{L}_{\rm CFF} = \sum_m \biggl| \mathcal{F}_q(t_m,\xi_m,Q_m^2) \\
	& & - \int_{-1}^{1} dx\, C^{+}(x,\xi_m)\, \hat{F}_q(x,\xi_m,t_m;Q_m^2) \biggr|^2, \nonumber
\end{eqnarray}
where $\hat{F}_q$ denotes the neural-network prediction for the corresponding GPD. Throughout this work, quantities with a hat indicate ML-inferred approximations to the associated physical observables.

This loss enforces agreement with experimental data only after integration over $x$, reflecting the fact that experiments constrain GPDs indirectly through CFFs rather than through pointwise information. Consequently, the learning problem is inherently nonlocal and underdetermined at the level of individual momentum fractions, underscoring the importance of combining CFF constraints with additional theoretical inputs. In practical implementations, the real and imaginary parts of the CFFs may be included separately in the loss function, with weights reflecting their experimental uncertainties and sensitivity to different regions of the GPD support. More broadly, the CFF loss anchors the ML model to experimentally accessible observables while encouraging the network to learn physically meaningful correlations across the full $x$ domain.

Notice that the CFF kernel, $C^+$, given in Eq..... is odd, so that only the odd GPD component, $H^+$ contributes. 

\subsection*{Mellin Moment Loss}
%\subsubsection{Mellin Moments}

In our framework, Mellin moments computed from \emph{ab initio} LQCD calculations can be incorporated as additional global constraints on the GPDs. 
The Mellin moments of GPDs defined in Eqs.\eqref{eq:H_moment} and \eqref{eq:E_moment}, are such that 
for odd $n$, the exponent of $x$ is even. Therefore, only the symmetric part of the GPD, namely, $H^-$, contributes to the integral.
Similarly, for even $n$, we only get a contribution from the antisymmetric part of the GPD ($H^+$).
%\begin{align*}
%    M_n^q &= \int_{-1}^{1}dxx^{n-1} \left(\frac{H^+{(x, \xi, t)} + H^-(x, \xi, t)}{2}\right)\\
%    &=\int_{0}^{1}dx x^{n-1}H^-(x, \xi ,t) \quad \quad \text{(odd $n$)}
%\end{align*}
%\begin{align*}
%    M_n^q = \int_{0}^{1}dx x^{n-1}H^+(x, \xi, t) \quad \quad \text{(even $n$)}
%\end{align*}
The results can be summarized as follows:
\begin{align*}
M_n &= 
\left\{
    \begin{array}{lr}
         \int_{0}^{1}dx x^{n-1}H^+(x, \xi, t) \quad \quad \text{(even $n$)}\\
         \\
         \int_{0}^{1}dx x^{n-1}H^-(x, \xi ,t) \quad \quad \text{(odd $n$)}
    \end{array}
\right.
\end{align*}
Therefore, we define the following loss functions,
\begin{eqnarray}
	\label{eq:L_MM}
	& & \mathcal{L}_{\rm MM}^+ =  \sum_{n=2,even}^{N_M-1} \sum_m \biggl| M_n^q(t_m,\xi_m,Q_m^2) \\
	& & \qquad - \int_{-1}^{1} dx\,	x^{n-1}\,	\hat{F}_q(x,\xi_m,t_m;Q_m^2)	\biggr|^2, \nonumber \\
	& & \mathcal{L}_{\rm MM}^- =  \sum_{n=1,odd}^{N_M} \sum_m \biggl| M_n^q(t_m,\xi_m,Q_m^2) \\
	& & \qquad - \int_{-1}^{1} dx\,	x^{n-1}\,	\hat{F}_q(x,\xi_m,t_m;Q_m^2)	\biggr|^2, \nonumber \\
    \end{eqnarray}
where $N_M$ denotes the highest Mellin moment included in the training. Together, the CFF and Mellin-moment losses provide complementary constraints: the former encode experimental information through nonlocal convolution integrals, while the latter enforce normalization and polynomiality properties rooted in local QCD operators.

%In the asymmetric frame, the formulae become:
%\begin{align*}
%M_n &= 
%\left\{
%    \begin{array}{lr}
%         \int_{\zeta/2}^{1}\frac{dX}{1-\zeta/2} \left(\frac{X-\zeta/2}{1-\zeta/2}\right)^{n-1}H^+(X, \zeta, t) \quad \quad \text{(even $n$)}\\
%         \\
%         \int_{\zeta/2}^{1}\frac{dX}{1-\zeta/2} \left(\frac{X-\zeta/2}{1-\zeta/2}\right)^{n-1}H^-(X, \zeta, t) \quad \quad \text{(odd $n$)}
%    \end{array}
%\right\}
%\end{align*}

\subsection*{Symmetry/Endpoint Loss}
In practice, the symmetry constraints can be incorporated either architecturally by: 1) restricting the NN prediction to the domain $x\in[0,1]$, and 2) reconstructing the negative -$x$ region using the exact symmetry relations — or through explicit penalty terms in the loss function. For vector GPDs, antisymmetry further implies $H_q^+(x=0,\xi,t)=0$, which can be enforced via the structural loss,
\begin{eqnarray}
	\label{eq:L_sym}
%	\mathcal{L}_{\rm sym} = \sum_m \left( \left| \hat{H}_q^+(0, \xi_m, t_m; Q_m^2) \right|^2 + \left| \hat{E}_q^+(0, \xi_m, t_m; Q_m^2) \right|^2 \right), \nonumber \\
	\mathcal{L}_{\rm sym} = \sum_m \left| \hat{H}_q^+(0, \xi_m, t_m; Q_m^2) \right|^2 , 
\end{eqnarray}
Similarly, the endpoint conditions are imposed through
\begin{eqnarray}
	\label{eq:L_end-point}
	\mathcal{L}_{\scriptsize \mbox{end-point}} = \sum_m \left| \hat{F}_q(1, \xi_m, t_m; Q_m^2) \right|^2.
\end{eqnarray}
Together with the CFF and Mellin-moment losses, these symmetry and boundary constraints complete the physics-informed loss structure, ensuring that the learned GPDs satisfy both global integral relations and exact pointwise properties dictated by QCD.

%%%%%
\subsection*{Bayesian Uncertainty Quantification}
%%%%%
%%%%%
{A key question in inverse problems is how the residual ambiguity in the reconstruction, after all available data and constraints are implemented, translates into uncertainties in the extracted functions. To capture this, we replace the deterministic neural network with a Bayesian neural network (BNN)\cite{mackay1992practical, neal2012bayesian}, where each weight and bias is modeled as a Gaussian random variable with trainable mean and standard deviation. The loss function then includes an additional Kullback–Leibler (KL) divergence term\cite{kullback1951information}.
\begin{equation}
    \mathcal{L}_{\rm KL} = D_{\rm KL}\big(q_{\bm\theta}(\mathbf{W}) \,\|\, p(\mathbf{W})\big),
\end{equation}
measuring the discrepancy between the learned variational posterior $q_{\bm\theta}$ and a standard-normal prior $p(\mathbf{W})$, which prevents the posterior from collapsing to a delta function and maintains a distribution over network configurations compatible with the data.}

{
During training, we draw a single set of weights from the current posterior at each forward pass using the reparameterization trick, allowing gradient-based optimization to proceed as usual. At inference time, we sample the posterior $N_s$ times, with each draw producing a different GPD prediction. The ensemble mean is taken as the central value, while the spread gives a pointwise estimate of the epistemic uncertainty, {\it i.e.}, how much freedom remains in the local shape of the GPD after imposing the integral constraints. In kinematic regions where the Mellin-moment and CFF constraints are strong, the posterior is narrow; in less constrained regions, it broadens, reflecting the limited information available in the training data.}

%\textcolor{red}{
{When the training data themselves carry statistical noise, the BNN posterior width incorporates both the epistemic contribution from the underdetermined inverse problem and the aleatoric contribution from fluctuations in the input data. The relative importance of the two sources can be separated by comparing predictions trained on noise-free and noise-injected pseudo-data (Sec.~\ref{sec:results}).
}
\section{Results}

\label{sec:results}

In the absence of sufficiently precise and abundant experimental determinations of the CFFs, to be used conjointly with the comprehensive LQCD Mellin moments results over a broad kinematic range, the goal of the present study is to perform a closure test, serving as a proof-of-principle demonstration of the proposed neural-network framework. 
Specifically, we assess whether a single neural-network representation of GPDs can simultaneously reproduce multiple classes of global integral constraints when trained on pseudo-data generated from a suitable phenomenological GPD model. At this stage we use the flexible parametrization from Refs.\cite{Kriesten:2021sqc,Panjsheeri:2025vpa} which obeys constraints from both deep inelastic scattering and elastic form factors experimental data and is therefore particularly apt to simulate data behavior. 
This controlled setting allows us to validate the internal consistency, expressivity, and stability of the NNGPD approach before applying it to real experimental and lattice inputs.

\subsection*{NN Architecture}
All results reported in this section are obtained using a fixed neural-network architecture, denoted as 
\[ {\rm NNGPD}(4,1,100) \] 
%%%%%
which is as a fully connected feedforward neural network that takes four kinematic variables $(x,\xi,t,Q^2)$, as input, and outputs a single generalized parton distribution. 
Each hidden layer has a width of 100 neurons. The same network architecture is used for all runs presented in this work, ensuring that the closure test probes the optimization and constraint structure of the framework rather than architectural flexibility.

The network employs the SiLU activation function at all hidden layers. To improve training stability and mitigate overfitting in this pseudo-data setting, layer normalization is applied after each linear transformation, and dropout with probability $p=0.1$ is used during training. No architectural or regularization hyperparameters are tuned between runs. Instead, all runs are performed with identical settings and differ only by random initialization of the network parameters.

\begin{figure*}
\centering
\includegraphics[width=0.75\linewidth]{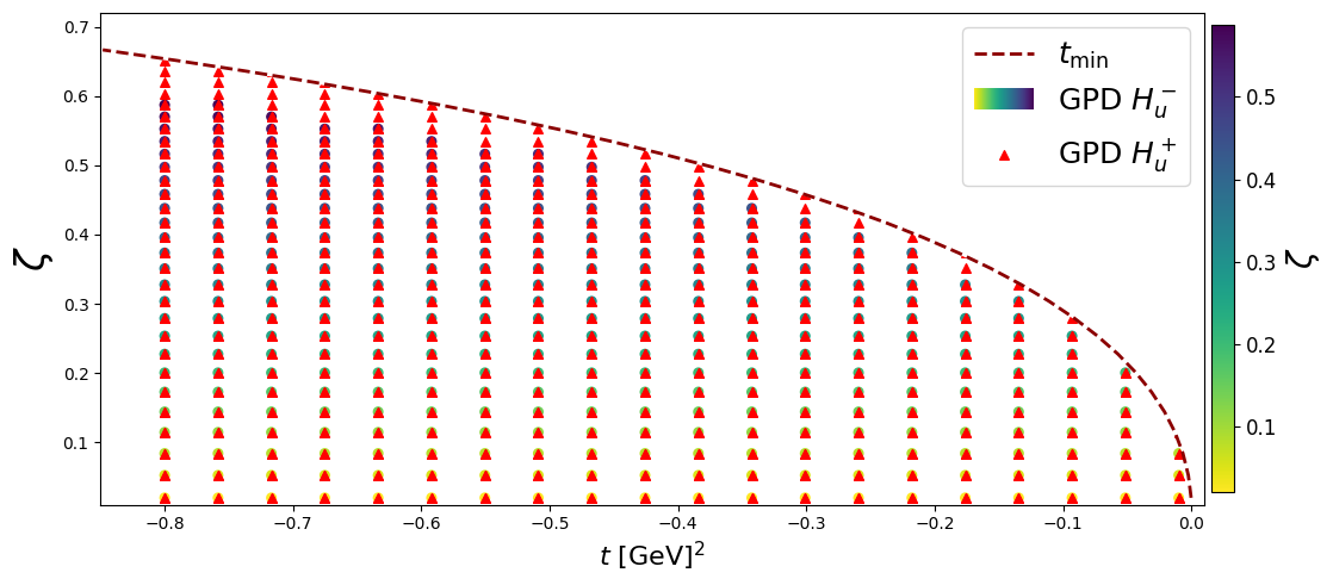}
\caption{Kinematic coverage of training dataset for the $H^-_u$ and $H^+_u$ distributions, shown at a  fixed value of $Q^{2}$. 
%The $t_{min}$ is calculated at $Q^2 = 6$ $\mathrm{GeV^2}$. 
}
    \label{fig:kinematic_plot}
\end{figure*}

\subsection*{Training data}
\begin{figure*}
\centering
\includegraphics[width=1.99\columnwidth]{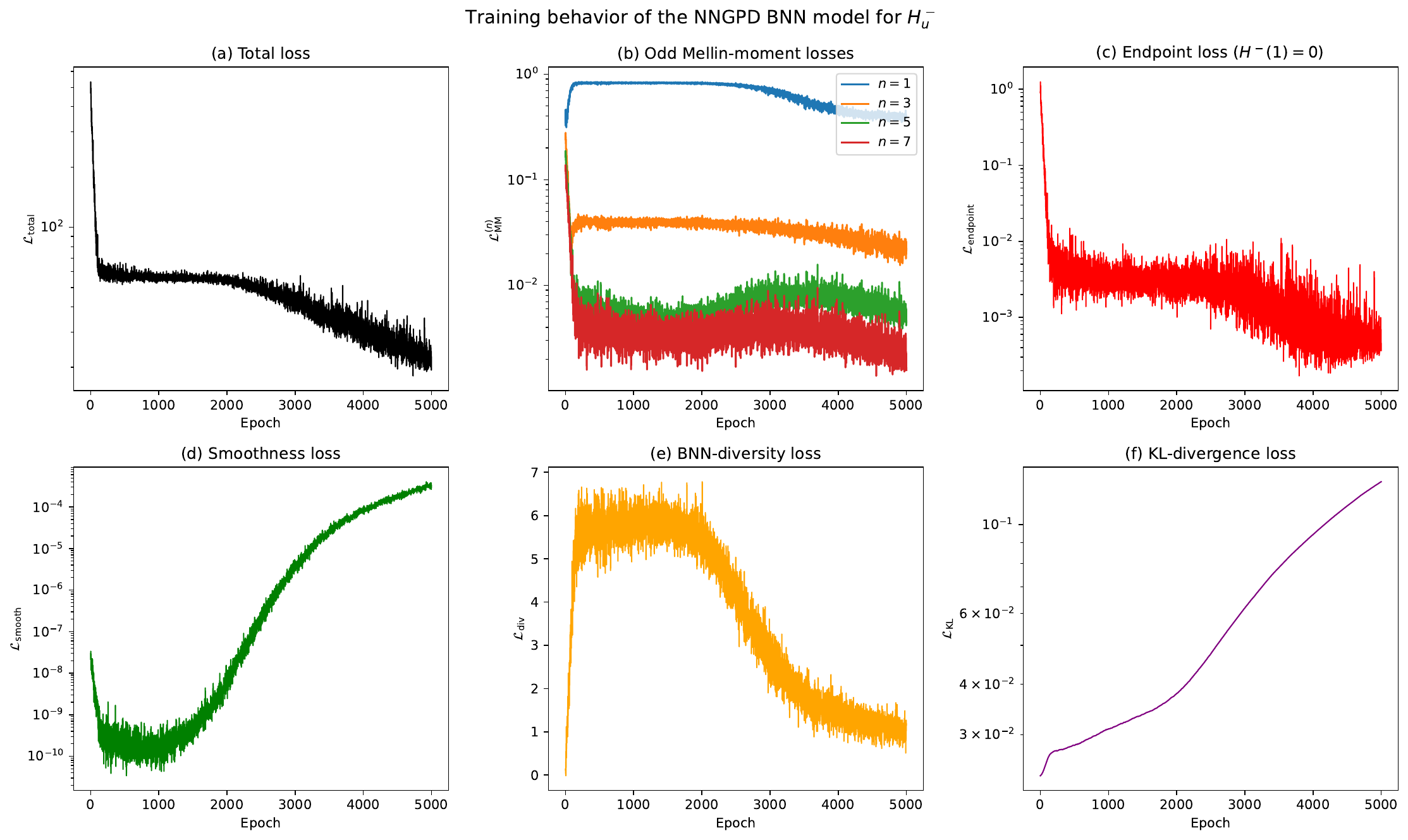}
\caption{Training behavior of the NNGPD BNN model for the symmetric combination $H^-_u$. (a) Total loss as a function of training epoch. (b) Losses associated with the odd Mellin-moment constraints ($n=1,3,5,7$). (c) Endpoint loss enforcing $H^-_q(1,\xi,t)=0$. (d) Smoothness loss $\mathcal{L}_{\rm smooth}$ penalizing large local variations of the learned GPD in $x$. (e) BNN-diversity loss $\mathcal{L}_{\rm div}$ maintaining posterior spread across the ensemble. (f) KL-divergence loss.}
\label{fig:training-hminus}
\end{figure*}

The training dataset consists of 9455 distinct kinematic points generated from the underlying phenomenological GPD model. Each data point corresponds to a unique $(Q^2,\xi,t)$ configuration and provides simultaneous pseudo-observables corresponding to Mellin moments up to $n=7$, and CFFs. The kinematic coverage of the dataset spans
\begin{eqnarray*}
Q^2 &\in& [4.0,\,10.0] \: \mathrm{GeV}^{2}, \quad
\xi \in [0.01,\,0.50], \\
\quad
t  &\in& -[0.01,\,0.80] \: \mathrm{GeV}^{2},
\end{eqnarray*}
which is representative of the domain relevant to existing and forthcoming DVCS measurements. {\footnote{Note that for every $\xi$, up to target mass corrections $t_{min} = -4 \xi^{2} / (1-\xi^{2})$. Any value of $t$ below this is kinematically forbidden.}} For each kinematic configuration, the network is constrained by the $n=2,4,6$ 
%second, fourth, and sixth 
Mellin moments, as well as by the real and imaginary parts of the corresponding Compton form factors. These quantities are treated on equal footing as global integral constraints derived from the same underlying GPD.
\begin{figure*}
\centering
\includegraphics[width=1.99\columnwidth]{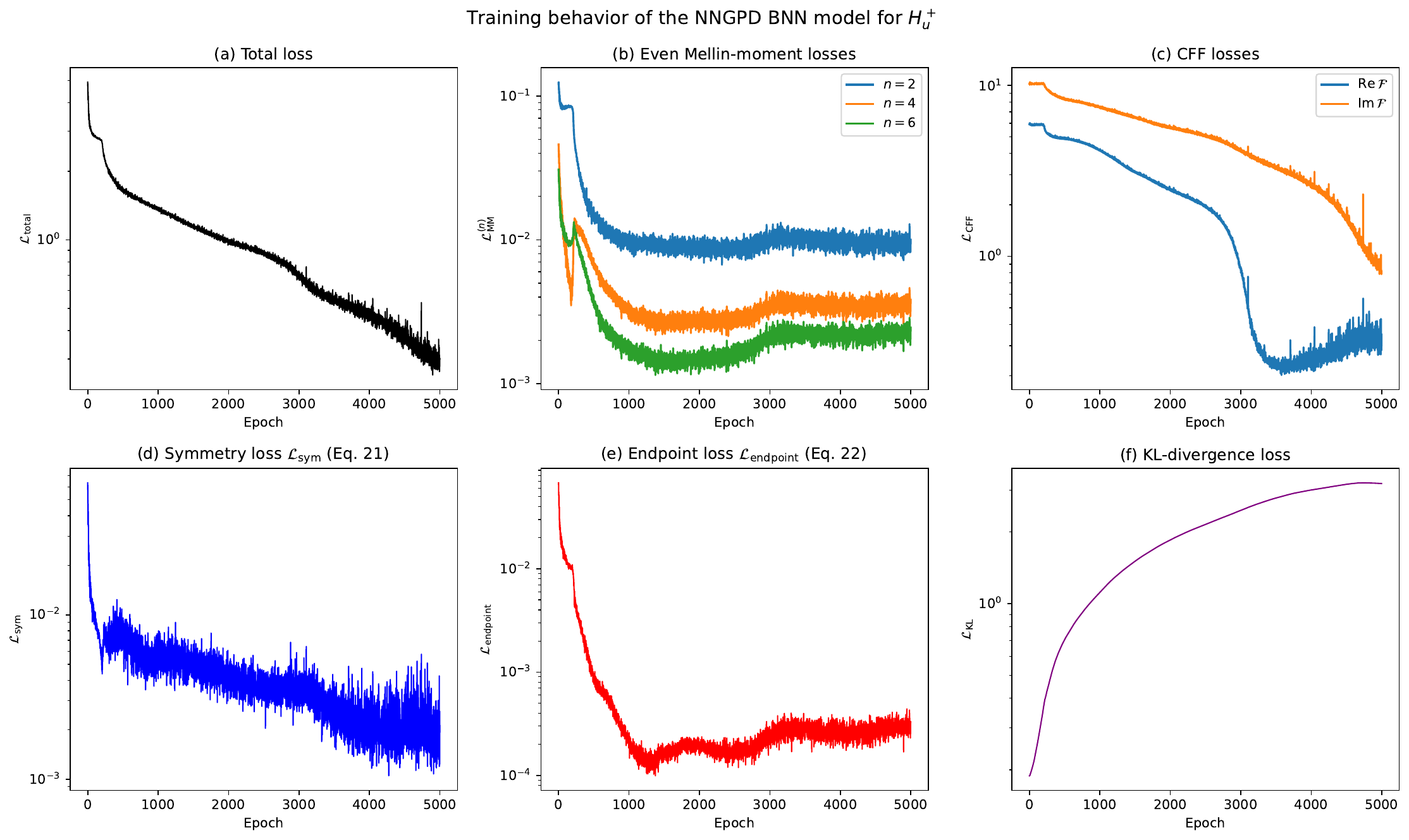}
%\caption{{Training behavior of the NNGPD BNN model. (a) Total loss as a function of training epoch. (b) Decomposition of the total loss into individual (unweighted) contributions. (c) Losses associated with even Mellin-moment constraints ($n=2,4,6$). (d) Endpoint loss enforcing the boundary conditions $H^+(0)=0$ and $H^+(1)=0$. (e) Losses for the real and imaginary parts of the Compton form factors. (f) KL-divergence loss measuring the departure of the BNN posterior from its prior.}}
\caption{{Training behavior of the NNGPD BNN model for the antisymmetric combination $H^+_u$. (a) Total loss as a function of training epoch. (b) Losses associated with the even Mellin-moment constraints ($n=2,4,6$). (c) Losses for the real and imaginary parts of the Compton form factors. (d) Symmetry loss $\mathcal{L}_{\rm sym}$ [Eq.~(\ref{eq:L_sym})] enforcing $H^+_q(0,\xi,t)=0$. (e) Endpoint loss $\mathcal{L}_{\rm end-point}$ [Eq.~(\ref{eq:L_end-point})] enforcing $H^+_q(1,\xi,t)=0$. (f) KL-divergence loss measuring the departure of the BNN posterior from its prior.}}
    \label{fig:training-hplus}
\end{figure*}
{
The optimization settings used to produce the figures in this section are broadly comparable. For the antisymmetric combination $H^+_u$, the curves in Fig.~\ref{fig:training-hplus} are obtained with AdamW~\cite{loshchilov2017decoupled} combined with Lookahead~\cite{zhang2019lookahead} and a ReduceLROnPlateau scheduler. For the symmetric combination $H^-_u$, a comparable adaptive optimization setup is used for the run shown in Fig.~\ref{fig:training-hminus}. In both cases, the training uses a batch size of 256. These settings correspond to the training runs shown in the figures below.
}

\begin{figure*}
\centering
\includegraphics[width=1.99\columnwidth]{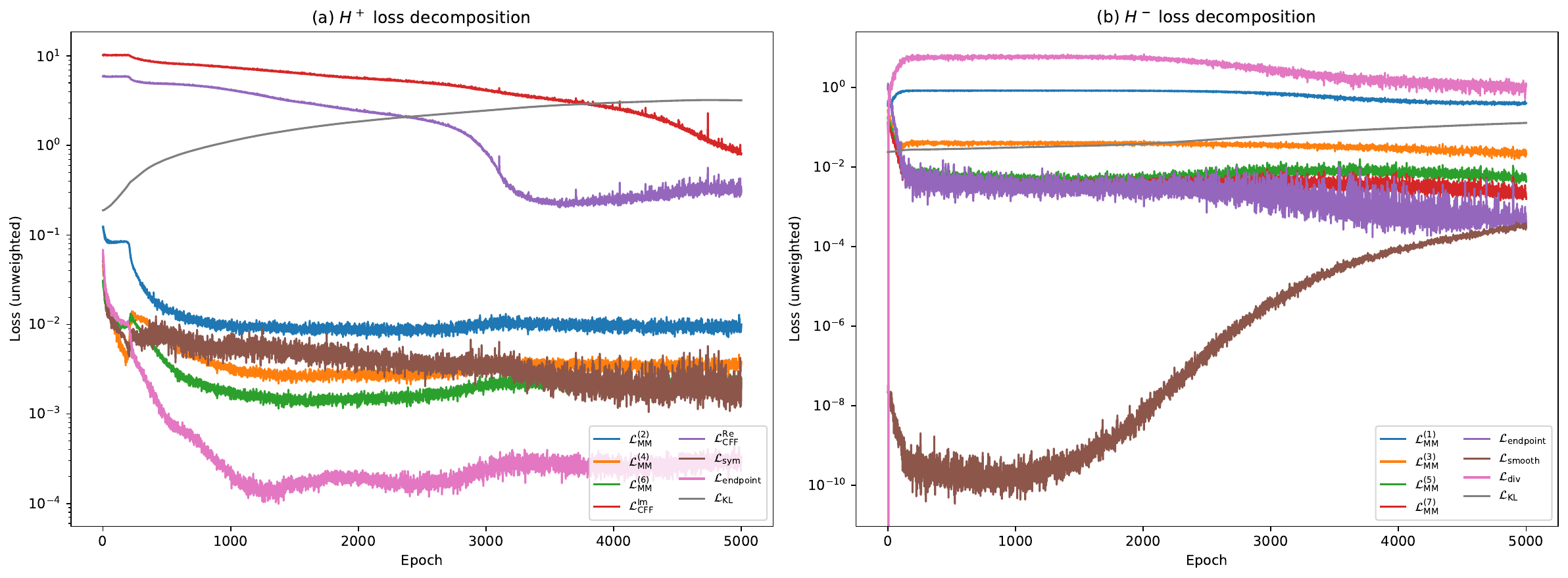}
\caption{Decomposition of the total loss into its individual (unweighted) contributions during training. Left panel: $H^+$ training (MM$_2$, MM$_4$, MM$_6$, Re/Im CFF, $\mathcal{L}_{\rm sym}$, $\mathcal{L}_{\rm end-point}$, KL). Right panel: $H^-$ training (MM$_1$, MM$_3$, MM$_5$, MM$_7$, endpoint, smoothness, diversity, KL).}
\label{fig:training-decomp}
\end{figure*}

\subsection*{Loss Function}
The total loss function minimized during training is constructed as a weighted sum of physically motivated contributions discussed in the previous section:
\begin{subequations}
\begin{eqnarray}
	\mathcal{L}_{\rm total}^+ & = &  w_{CFF} \mathcal{L}_{CFF} + w_{MM}^+ \mathcal{L}_{MM}^+  + w_{EP}^+ \mathcal{L}_{EP}^+   \nonumber \\
 &   + & w_{KL} \mathcal{L}_{KL}
      \\
	\mathcal{L}_{\rm total}^- & = &  w_{MM}^- \mathcal{L}_{MM}^- + w_{EP}^- \mathcal{L}_{EP}^- + w_{\rm smooth} \mathcal{L}_{\rm smooth} \nonumber \\
 &   + & w_{\rm diversity} \mathcal{L}_{\rm div} +  w_{KL} \mathcal{L}_{KL}
\end{eqnarray}
\end{subequations}
{
Here ${\rm CFF,MM,EP,KL}$ refer to the loss-function contributions defined above, with $\mathcal{L}_{\rm CFF}$ given in Eq.~(\ref{eq:L_CFF}) and $\mathcal{L}_{\rm MM}$ in Eq.~(\ref{eq:L_MM}). The shorthand $\mathcal{L}_{EP}$ denotes the boundary contribution: for the antisymmetric sector, $\mathcal{L}_{EP}^+$ combines the symmetry term of Eq.~(\ref{eq:L_sym}) and the $x=1$ boundary term of Eq.~(\ref{eq:L_end-point}); for the symmetric sector, $\mathcal{L}_{EP}^-$ reduces to the $x=1$ boundary term alone.
}Note that $ \mathcal{L}_{CFF}$ includes loss terms associated with the real and imaginary parts of the CFFs, Eq.~(\ref{eq:L_CFF}); 
%which can be straightforwardly split from the definition of the MSE of the complex CFF in , a loss term $\mathcal{L}_{\rm MM}$ defined in Eq.~(\ref{eq:L_MM}) enforcing agreement with the Mellin moments,  
%\st{a symmetry-enforcing loss that imposes the required even or odd behavior of the GPDs under $x \to -x$, and an $L_2$ regularization term on the network parameters.} 
{
The Kullback--Leibler divergence term $\mathcal{L}_{\rm KL}$ regularizes the approximate posterior of the Bayesian neural network toward its prior.
} 
{an endpoint loss enforcing the boundary conditions $H^+_q(0)=0$ and $H^+_q(1)=0$, and a Kullback--Leibler divergence term $\mathcal{L}_{\rm KL}$ that regularizes the approximate posterior of the Bayesian neural network toward its prior.} 
The relative weights of the different contributions are chosen as
\begin{eqnarray*}
& & w_{\rm MM} = 10, \quad
w_{\rm CFF}^{\rm Re} = 0.1, \quad
w_{\rm CFF}^{\rm Im} = 0.1, \\
& & \qquad w_{\rm KL} = 0.001, \quad w_{\scriptsize \mbox{end-point}} = 10\end{eqnarray*}
%\st{with an $L_2$ regularization coefficient $\lambda_{\theta} = 5\times10^{-4}$ and the smoothness constraint $\lambda_x = 0$.}
{For the symmetric $H^-$ distribution, the weights differ from the $H^+$ case, reflecting both the different set of Mellin moments used ($n=1,3,5,7$) and the need to compensate for their rapidly decreasing absolute magnitudes with increasing $n$. We choose
\begin{equation*}
\begin{aligned}
w_{\mathrm{MM}_1} &= 10, \quad w_{\mathrm{MM}_3} = 160, \\
w_{\mathrm{MM}_5} &= 730, \quad w_{\mathrm{MM}_7} = 2500, \\
w_{\scriptsize \mbox{end-point}} &= 1, \quad w_{\rm smooth} = 0.5, \\
w_{\rm diversity} &= 5, \quad w_{\rm KL} = 0.001.
\end{aligned}
\end{equation*}
The $H^-$ training further includes the smoothness penalty
\begin{equation*}
\mathcal{L}_{\rm smooth} = \sum_m \int dx\, \left|\frac{\partial \hat H^-_q}{\partial x}\right|^2,
\end{equation*}
and the BNN-diversity penalty
\begin{equation*}
\mathcal{L}_{\rm div} = -\mathbb{E}\!\left[\log \mathrm{Var}(\hat H^-_q)\right],
\end{equation*}
weighted by $w_{\rm smooth}$ and $w_{\rm diversity}$, respectively. In practice, $\mathcal{L}_{\rm smooth}$ is implemented as a first-difference penalty on the predicted curve evaluated on the $x$ grid, which provides a discrete approximation to this derivative-based regularization. The diversity term prevents the posterior variance from collapsing to zero during training. These additional terms are needed because odd Mellin moments alone constrain $H^-$ much less tightly than the combination of even moments and Compton form factors constrains $H^+$. }These weights are fixed across all runs and are selected to balance the numerical scales of the pseudo-observables while emphasizing exact symmetry properties.

{The choice of loss weights is guided by two considerations: (i) the different numerical scales of individual loss contributions, and (ii) the need to enforce a hierarchy that reflects the physical importance of each constraint. The values of different terms span a wide range of numerical magnitudes. For example, the Mellin moments, being integrals of the GPD weighted by powers of $x$, are typically much smaller in absolute value than the Compton form factors. As a result, the raw (unweighted) loss contributions from the Mellin-moment terms would be suppressed relative to those from the CFF terms. The weight assignment $w_{\mathrm{MM}} \gg w_{\mathrm{CFF}}$ counteracts this imbalance and ensures that the optimizer allocates comparable gradient pressure to both classes of constraints.}

Figure~\ref{fig:training-hplus} summarizes the training dynamics of the NNGPD model for $H^+_u$ and the interplay among the different components of the total loss. Panel (a) shows the evolution of the total loss as a function of training epoch. After a rapid decrease during the early stages of training, the loss enters a slower relaxation regime and continues to decrease more gradually over the remainder of training, indicating a stable optimization trajectory. {A full decomposition of the total loss into its individual unweighted contributions is provided as a separate plot in Fig.~\ref{fig:training-decomp} (left panel), where the clear separation of time scales between data-driven losses and constraint-driven terms is visible.}

The running of the Mellin-moment loss with the epochs is shown in Fig.~\ref{fig:training-hplus}{(b)}. Following an initial sharp decrease, this loss relaxes more gradually and approaches a stable value, reflecting the nontrivial task of satisfying global integral constraints over the parton momentum fraction while simultaneously fitting Compton form factor (CFF) data. The smooth saturation indicates that the network reconciles these competing constraints without overfitting individual moments or kinematic points. Complementary information is provided by Fig.~\ref{fig:training-hplus}\textcolor{red}{(c)}, which displays the losses associated with the real and imaginary parts of the CFFs. Both components decrease monotonically with training, with the imaginary part showing a somewhat faster decrease over most of the training. The simultaneous convergence of both components demonstrates that the network captures the full complex structure of the amplitudes rather than fitting one component at the expense of the other.

{{Figures~\ref{fig:training-hplus}(d) and (e) show, respectively, the symmetry loss $\mathcal{L}_{\rm sym}$ enforcing $H^+(0)=0$ and the endpoint loss $\mathcal{L}_{\rm end-point}$ enforcing $H^+(1)=0$.} Both losses decrease rapidly at early epochs and continue to fall throughout training, indicating that the pointwise boundary constraints are steadily absorbed as the network refines the shape of the GPD near the kinematic boundaries. Figure~\ref{fig:training-hplus}(f) displays the KL-divergence loss, which measures the departure of the learned BNN posterior from its standard-normal prior. In contrast to the data-driven losses, this term increases with training epoch, reflecting the expected behavior of a Bayesian neural network: as the posterior sharpens around configurations that satisfy the physics constraints, it necessarily moves away from the prior. The modest overall magnitude of $\mathcal{L}_{\rm KL}$, combined with its small weight $w_{\rm KL} = 0.001$, confirms that the KL penalty acts as a mild regularizer without impeding the network's ability to learn the physically relevant structure.}

{The corresponding training dynamics for the symmetric combination $H^-_u$ are collected in Fig.~\ref{fig:training-hminus}. The total loss, panel (a), decreases rapidly during the first few hundred epochs and settles into a long plateau whose overall level remains higher than that of $H^+$, reflecting the fact that $H^-$ is constrained only by odd Mellin moments ($n=1,3,5,7$) and a boundary condition at $x=1$, with no CFF information to anchor the shape pointwise. Panel (b) shows that all four odd-moment losses decrease and stabilize. Among the unweighted contributions, $\mathcal{L}_{\mathrm{MM}}^{(1)}$ remains the largest, while the higher-order moments are progressively smaller in magnitude; this hierarchy motivates the use of increasingly larger weights for higher moments, so that their contributions to the total loss remain numerically comparable during optimization. Panel (c) confirms that the endpoint condition $H^-(1)=0$ is steadily absorbed. The smoothness and BNN-diversity losses [panels (d) and (e)] do not decrease monotonically: the former drifts upward by several orders of magnitude in the second half of training, while the latter departs from its initial minimum and broadens. Both behaviors are expected consequences of the weaker constraint structure available to $H^-$---the network must develop stronger local gradients in $x$ to reproduce the higher-order moments, and the posterior remains correspondingly broader across kinematic configurations. The KL-divergence loss, panel (f), grows throughout training, as in the $H^+$ case, mirroring the expected BNN dynamics in which the variational posterior moves away from its prior to accommodate the physics constraints.}

{Figure~\ref{fig:training-decomp} collects the unweighted loss components for both training runs on a common logarithmic axis, making explicit the widely different numerical scales involved. For $H^+$ (left panel) the CFF losses $\mathcal{L}_{\mathrm{CFF}}^{\mathrm{Re}}$ and $\mathcal{L}_{\mathrm{CFF}}^{\mathrm{Im}}$ dominate ($\sim 10^{0}$--$10^{1}$), the Mellin-moment losses settle at $\sim 10^{-2}$--$10^{-3}$, and the pointwise boundary losses $\mathcal{L}_{\rm sym}$ and $\mathcal{L}_{\rm end\text{-}point}$ reach $\sim 10^{-3}$--$10^{-4}$; this hierarchy is consistent with assigning larger relative weight to the Mellin-moment and endpoint terms than to the CFF terms in the total loss. For $H^-$ (right panel) $\mathcal{L}_{\mathrm{MM}}^{(1)}$ is the largest unweighted contribution ($\sim 10^{0}$), while the higher-order moments drop by roughly one order of magnitude per increment in $n$, reaching $\sim 10^{-2}$ for $\mathcal{L}_{\mathrm{MM}}^{(7)}$; this steep scaling is compensated by assigning progressively larger weights to the higher-order moments, so that the weighted moment terms contribute on more comparable numerical scales to the total loss during optimization. In both panels the KL-divergence contribution rises monotonically---consistent with the Bayesian interpretation where the variational posterior must depart from its prior to reproduce the physics constraints---while all data-driven and boundary losses relax downward.}
% \begin{figure}
% \centering
% \includegraphics[width=0.99\columnwidth]{benchmark.pdf}
% \caption{Benchmark comparison between the phenomenological spectator-based generalized parton distribution (GPD) and the NNGPD prediction. The dashed curves show the ``ground-truth'' GPD $H(x,\xi,t)$ from the underlying phenomenological model used to generate the pseudo-data, while the solid curves denote the ensemble-averaged neural-network prediction. The shaded bands indicate the $\pm 3\sigma$ uncertainty estimated from independent training runs. Vertical dotted lines mark the kinematic boundaries $x=\pm\xi$ separating the ERBL and DGLAP regions. The three panels correspond to different kinematic points $(Q^2,\xi,t)$, as indicated in each panel, and probe distinct regimes of skewness and momentum transfer.
% }
% \label{fig:benchmark}
% \end{figure}
{\color{red}
\begin{figure*}
\centering
\includegraphics[width=1.99\columnwidth]{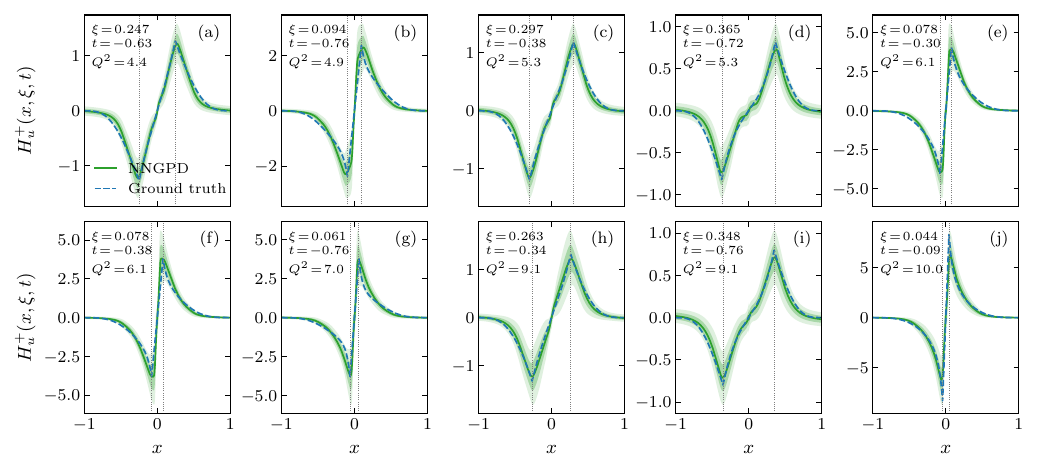}
\caption{{BNN predictions for the antisymmetric GPD combination $H^+_u(x,\xi,t)$ at ten representative kinematic configurations, labeled (a)--(j). The network is trained with even Mellin-moment constraints ($n=2,4,6$) and Compton form factor data. The solid green curves denote the NNGPD ensemble mean, the dashed blue curves show the ground-truth GPD from the underlying phenomenological model, and the shaded green bands indicate the $1\sigma$ and $2\sigma$ posterior uncertainty regions. The vertical dotted lines mark the kinematic boundaries $x = \pm\xi$ separating the DGLAP and ERBL regions.}}
\label{fig:hplus-bnn}
\end{figure*}
}

{
%We now present the BNN-based results, in which the antisymmetric and symmetric GPD combinations are reconstructed separately. 
Figure~\ref{fig:hplus-bnn} displays the BNN predictions for $H^+_u(x,\xi,t)$ at ten kinematic configurations of relevance for the programs at Jefferson Lab and EIC. This combination is constrained by even Mellin moments ($n=2,4,6$) together with the real and imaginary parts of the Compton form factors; the latter contribute because the antisymmetric Wilson coefficient $C^-(x,\xi)$ in Eq.~(\ref{eq:WCF}) projects exclusively onto $H^+$. The BNN ensemble mean (green) tracks the ground-truth GPD (dashed blue) closely in every panel, reproducing both the antisymmetric sign change through $x=0$ and the cusp-like transitions at $x = \pm\xi$. The posterior uncertainty bands are narrow, at the level of a few percent of the peak amplitude, confirming that the combined CFF and even-moment data determine $H^+$ with good precision. Minor deviations between the mean prediction and the ground truth are visible near $x = \pm\xi$, where the GPD varies most rapidly, but they remain well within the $2\sigma$ band.
}

{\color{red}
\begin{figure*}
\centering
\includegraphics[width=1.99\columnwidth]{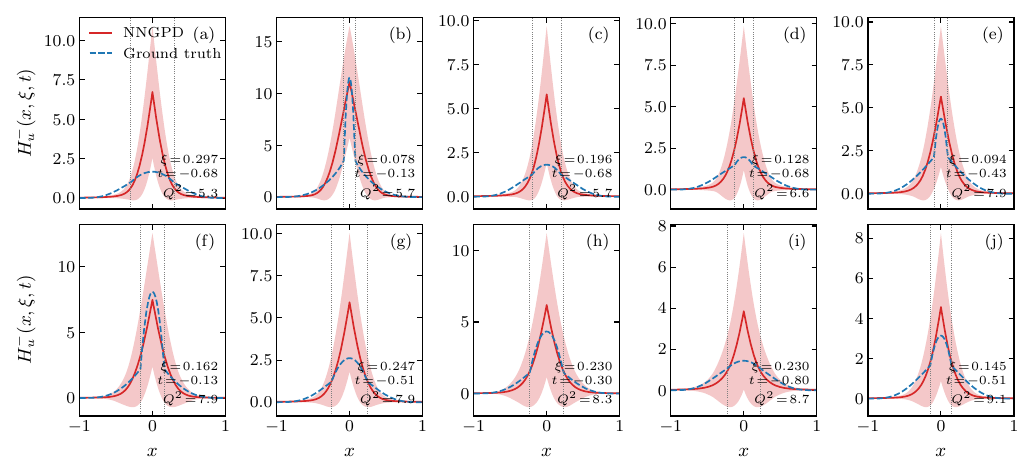}
\caption{{BNN predictions for the symmetric GPD combination $H^-_u(x,\xi,t)$ at ten representative kinematic configurations, labeled (a)--(j). The network is trained with odd Mellin-moment constraints ($n=1,3,5,7$) only. The solid red curves denote the NNGPD ensemble mean, the dashed blue curves show the ground-truth GPD from the underlying phenomenological model, and the shaded red band indicates the $1\sigma$ posterior uncertainty region. The vertical dotted lines mark the kinematic boundaries $x = \pm\xi$. The uncertainty bands are significantly wider than for $H^+$ (cf.\ Fig.~\ref{fig:hplus-bnn}), reflecting the weaker constraining power of odd Mellin moments alone.}}
\label{fig:hminus-bnn}
\end{figure*}
}
{
The situation is quite different for the symmetric combination $H^-_u$, shown in Fig.~\ref{fig:hminus-bnn}. Since $H^-$ is symmetric under $x \to -x$, it does not contribute to the CFF convolution and is therefore constrained only by odd Mellin moments ($n=1,3,5,7$). As a result, the posterior uncertainty bands are roughly an order of magnitude wider than for $H^+$. The BNN mean (red) also shows noticeable deviations from the ground truth in several kinematic configurations.}

{While the network captures the overall shape—a positive, roughly bell-shaped distribution peaked near $x = 0$—its accuracy worsens at larger $|t|$, where the underlying model becomes more structured. This difference reflects a limitation of the available constraints: odd Mellin moments alone, even up to $n=7$, do not provide enough information to fully determine the detailed $x$-dependence of $H^-$. Improving the reconstruction would likely require observables that are directly sensitive to the symmetric sector, such as charge-separated DVCS cross sections or pseudoscalar meson production amplitudes.
}
{
\begin{figure*}
\centering
\includegraphics[width=1.99\columnwidth]{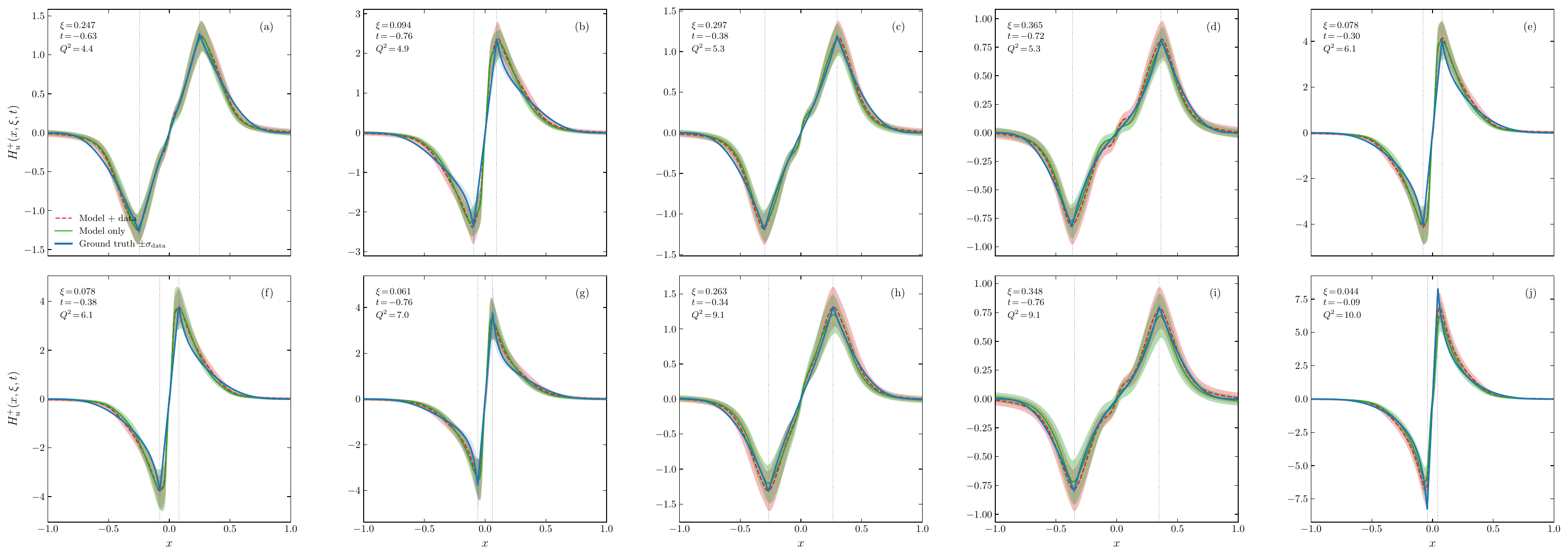}
\caption{{Comparison of model-only and model-plus-data uncertainty estimates for $H^+_u(x,\xi,t)$ at ten representative kinematic configurations, labeled (a)--(j). The green bands show the $1\sigma$ posterior uncertainty from a BNN trained on noise-free pseudo-data (epistemic uncertainty only), while the red bands show the $1\sigma$ uncertainty from a BNN trained with Gaussian noise injected into the training targets (epistemic plus aleatoric uncertainty).  The vertical dotted lines mark the kinematic boundaries $x = \pm\xi$.}}
\label{fig:errorbar-comparison}
\end{figure*}
}

\begin{figure*}
\centering
\includegraphics[width=7cm]{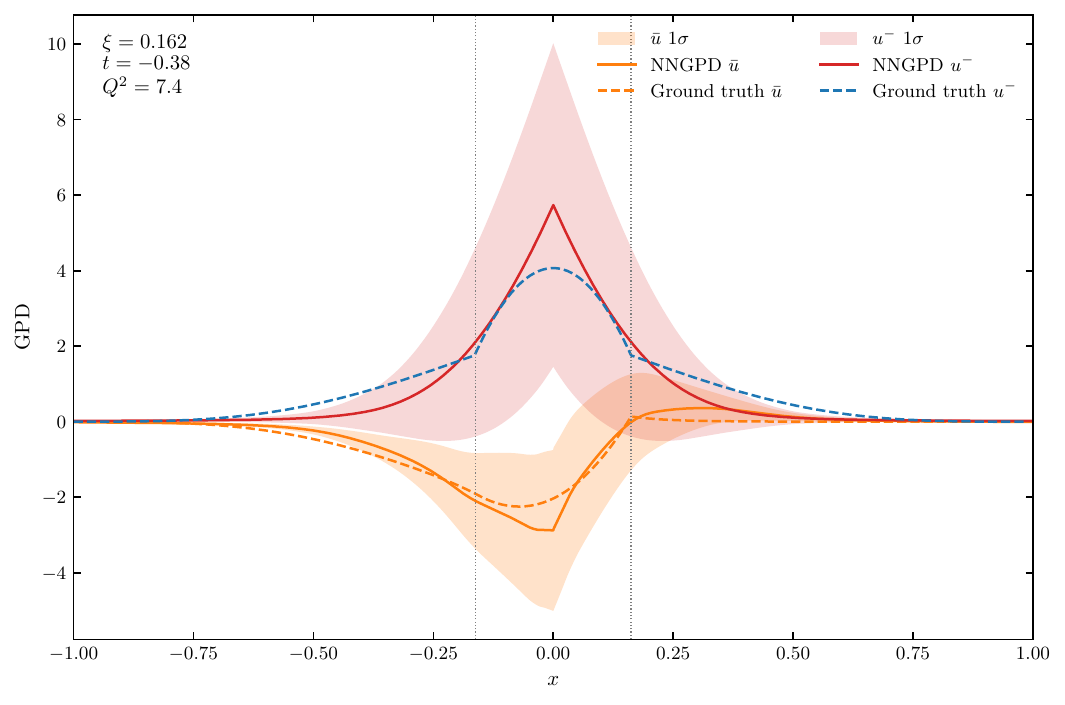}
\caption{Example of a comparison of  the valence, $H_{u_v}(x,\xi,t)$, and sea quark, $H_{\bar{u}}(x,\xi,t)$, GPDs, for the kinematics: $\xi= 0.162$, $t=-0.38$ GeV$^2$, $Q^2=7.4$ GeV$^2$, chosen in a range compatible with Jefferson Lab measurements.  Notation as in Fig.\ref{fig:errorbar-comparison}.}
\label{fig:GPD-comparison}
\end{figure*}

{
Figure~\ref{fig:errorbar-comparison} compares the uncertainty in the $H^+$ reconstruction under two training setups: one using noise-free pseudo-data (green bands, representing epistemic uncertainty only), and another with Gaussian noise added to the training targets (red bands, capturing both epistemic and aleatoric components). Across all ten kinematic points, the two bands largely overlap, with the noise-injected case showing only a slight broadening. This suggests that the dominant source of uncertainty comes from the underdetermined nature of the inverse problem, rather than from data noise. This behavior is expected, as the integral constraints tend to average over fluctuations in the inputs, leaving the remaining ambiguity mainly driven by the non-uniqueness of the GPD reconstruction. In future studies using real experimental data—particularly in kinematic regions with limited statistics—the contribution from aleatoric uncertainty may become more significant.
}

In Figure \ref{fig:GPD-comparison} we show the valence and sea quark GPDs, $H_{u_v} = H_u^-$, and $H_{\bar{u}}$, Eqs.\eqref{eq:Hplusminus}, respectively. The dashed line (ground truth) corresponds to the parametrization of Ref.\cite{Panjsheeri:2025vpa}. 
{
\begin{figure*}
\centering
\includegraphics[width=1.99\columnwidth]{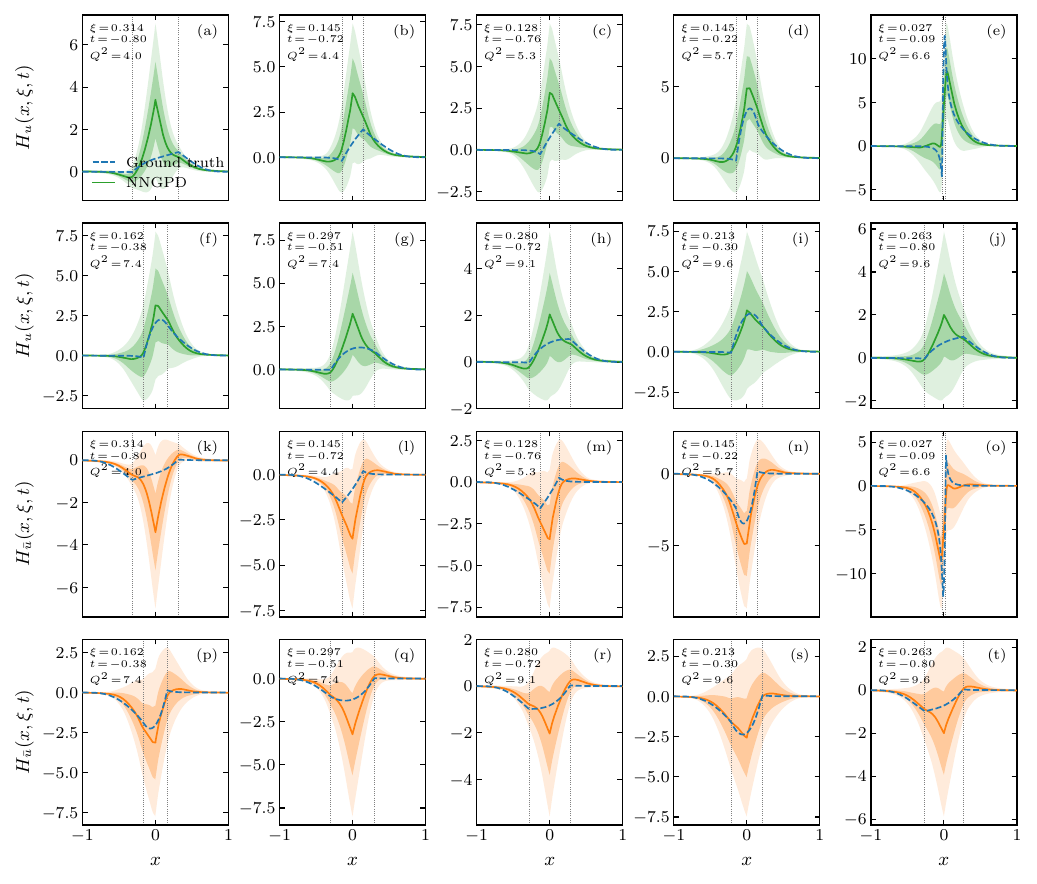}
\caption{{Quark and antiquark GPDs reconstructed from the BNN predictions of $H^+$ and $H^-$, at ten representative kinematic configurations. Panels (a)--(j) show the quark distribution $H_u(x,\xi,t) = (H^+ + H^-)/2$ (green bands) and panels (k)--(t) show the antiquark distribution $H_{\bar{u}}(x,\xi,t) = (H^+ - H^-)/2$ (orange bands). The solid colored curves denote the NNGPD ensemble mean, the dashed blue curves show the ground-truth GPD from the underlying phenomenological model, and the shaded bands indicate the $1\sigma$ and $2\sigma$ posterior uncertainty regions. The vertical dotted lines mark the kinematic boundaries $x = \pm\xi$.}}
\label{fig:hq-hqbar-bnn}
\end{figure*}
}
{One can see that, within our framework, one can constrain the $H^{+}$ distribution to high precision, due to the constraint from the CFFs, while the $H^{-}$ distribution emerges imprecisely. The consequence is that the resolution on the separate quark flavor GPDs is not as accurate. 
%which has consequences for QCD evolution. 
On the other hand, the angular momentum sum rule is given by the $H^{+}$ component and one can therefore obtain a precise estimate for the nucleon angular momentum, by adding the constraint of polynomiality.} In Figure \ref{fig:hq-hqbar-bnn} the $u_v/u_{\bar{u}}$ separated GPDs are shown for all of the additional kinematics used in this work. \\

Taken together, these benchmark results demonstrate that the NNGPD framework is capable of reconstructing a physically consistent GPD from a limited and heterogeneous set of integral constraints, without relying on direct pointwise supervision. Even in this controlled closure-test setting, the task is highly nontrivial: the neural network must simultaneously satisfy Mellin-moment constraints, convolution relations defining Compton form factors, exact symmetry requirements, and regularization, each probing different projections of the underlying function. The observed level of agreement confirms that the constrained-learning strategy successfully encodes these competing requirements into a single, globally consistent functional representation.

While the agreement is not exact—reflecting both the indirect nature of the constraints and the finite expressive capacity enforced by regularization—the deviations are controlled and quantitatively understood. This level of performance is encouraging for future applications, where experimental uncertainties, limited kinematic coverage, and lattice-QCD systematics will further complicate the inverse problem. The present benchmark results therefore provide a strong validation of the NNGPD approach and motivate its extension to realistic global analyses combining experimental and lattice-QCD inputs.

\section{Conclusion and Outlook}

\label{sec:conclusion}

In this work, we have developed a neural-network framework, referred to as NNGPD, for representing GPDs in a manner constrained directly by physically motivated integral relations. Rather than assuming a specific functional ansatz for GPDs, the framework treats their determination as a global inverse problem, in which experimental and theoretical information enters through loss functions enforcing convolution integrals for Compton form factors, Mellin moments associated with generalized form factors, and exact symmetry and boundary conditions dictated by QCD. This formulation combines the expressive flexibility of neural networks with the rigorous theoretical structure underlying GPD phenomenology.

As a proof-of-principle demonstration, we performed a closure test using pseudo-data generated from a phenomenological spectator-based GPD model. The neural network was trained exclusively on aggregate observables—namely Compton form factors and Mellin moments—without any direct pointwise supervision in the momentum fraction $x$. Despite this highly indirect training setup, the resulting NNGPD reconstruction reproduces the main features of the underlying model GPDs across a broad kinematic domain. The observed training dynamics, stability across independent runs, and benchmark comparisons demonstrate that the constrained-learning strategy is capable of encoding multiple, competing integral constraints into a single, globally consistent functional representation.

From a methodological standpoint, the present approach differs fundamentally from conventional phenomenological parametrizations of GPDs, based on fixed functional ans\"{a}tze such as double-distribution constructions or conformal partial-wave expansions, characterized by a limited number of shape parameters. While these frameworks incorporate important theoretical constraints by construction, they inevitably introduce model-dependent assumptions about functional form and kinematic correlations. By contrast, the neural-network representation employed here is deliberately agnostic about the detailed functional structure of GPDs, 
%Physical information enters exclusively through experimentally and theoretically established integral relations, including Compton form factors and Mellin moments. 
since the role traditionally played by a chosen parametrization is replaced by a flexible, data-driven function space constrained only by global physical observables. This perspective aligns naturally with the inverse-problem character of GPD phenomenology and provides a systematic avenue for incorporating all available inputs -- from experimental measurements to LQCD calculations -- within a unified framework including a robust treatment of uncertainty quantification through the use of BNN.

%  and  using a data-driven, 
Constructing GPDs from experimental data with statistically faithful uncertainties, and theory symmetry inputs with minimal parametrization bias -- in conjunction with LQCD -- takes on renewed urgency as new data from Jefferson lab are made available, and in light of the upcoming EIC, 
%The EIC will provide unprecedented luminosity, polarization control, and kinematic reach, enabling precision measurements of exclusive processes over a wide range of $x$, $\xi$, and $t$. This 
whose experimental program is expected to dramatically improve constraints on quark and gluon GPDs, including their flavor, spin, and spatial dependence, and to extend such studies to nuclei for the first time. In this context, neural-network representations of GPDs provide a natural and powerful avenue.
%The anticipated volume and precision of EIC data pose both an opportunity and a challenge for GPD phenomenology 
%On the one hand, they promise transformative advances in our understanding of nucleon and nuclear structure.
%that demands analysis frameworks which are sufficiently flexible to accommodate high-dimensional functional dependence, yet sufficiently constrained to ensure physical consistency and interpretability. 
%offering the ability to learn complex functional forms directly from data while incorporating theoretical constraints through architecture design, loss functions, or training strategies.

%Motivated by these considerations, the present work develops a neural-network approach to representing GPDs based on Mellin moments and CFF, with the goal of providing a scalable and robust framework for future global analyses. Such approaches are particularly well suited for the EIC era, where the combination of high-statistics data and multidimensional observables will require novel strategies for efficient, accurate, and unbiased GPD reconstruction.

Looking ahead, a natural next step is to extend this framework beyond closure tests to applications involving real experimental and lattice-QCD inputs. On the experimental side, this includes incorporating Compton form factors extracted from various DVES measurements, along with realistic uncertainties and correlations. 
%as well as data anticipated from the forthcoming Electron--Ion Collider. 
On the theoretical side, LQCD calculations of Mellin moments can be integrated in a systematic manner using the present framework where all different inputs enter on equal footing through global loss terms.

Several methodological improvements and extensions also warrant further investigation. One important direction concerns the implementation of polynomiality constraints, which require that Mellin moments of GPDs be polynomials in the skewness parameter $\xi$ with coefficients given by generalized form factors. 
While polynomiality is partially enforced in the present study through the inclusion of low-order Mellin moments, future work could incorporate it directly, either by introducing dedicated loss terms that penalize violations of polynomial structure or by designing neural-network architectures that encode polynomiality by construction. Such developments would further restrict the hypothesis space to physically admissible GPDs and improve extrapolation into regions that are only weakly constrained by available data.
%%% 
%LQCD Mellin moments calculations are not avaliable at present for $\xi\neq 0$, however.
%%%%

Furthermore, the NNGPD framework allows for the inclusion of additional theoretical constraints as they become relevant. These may include positivity bounds, dispersion-relation constraints linking the real and imaginary parts of Compton form factors, kinematic higher twists and beyond LO PQCD evolution terms. From a machine-learning perspective, future studies may also explore alternative architectures, such as equivariant or operator-learning models, as well as adaptive strategies for balancing competing loss terms when confronting real data with nonuniform uncertainties.

In summary, the present study establishes a viable and systematically improvable pathway for extracting generalized parton distributions using constrained machine learning. By demonstrating that a neural network can faithfully reconstruct GPDs from global integral information alone, this work lays the foundation for future phenomenological applications that combine experimental measurements and lattice-QCD calculations within a unified, data-driven framework.

\acknowledgements
This work was done under DOE grants DE-SC0016286, DE-SC0024644.

\bibliography{references}

\end{document}